\documentclass[fleqn,10pt]{wlscirep}

\usepackage{url}
\usepackage{bbm}

\usepackage{color}

\title{{Metrics for network comparison using egonet feature distributions}}

\author[1,*]{Carlo Piccardi}

\affil[1]{Department of Electronics, Information and Bioengineering, Politecnico di Milano, Piazza Leonardo da Vinci 32, 20133 Milano, Italy}
\affil[*]{carlo.piccardi@polimi.it}

\begin{abstract}
\textbf{Identifying networks with similar characteristics in a given ensemble, or detecting pattern discontinuities in a temporal sequence of networks, are two examples of tasks that require an effective metric capable of quantifying network (dis)similarity. Here we propose a method based on a global portrait of graph properties built by processing local nodes features. More precisely, a set of dissimilarity measures is defined by elaborating the distributions, over the network, of a few egonet features, namely the degree, the clustering coefficient, and the egonet persistence. The method, which does not require the alignment of the two networks being compared, exploits the statistics of the three features to define one- or multi-dimensional distribution functions, which are then compared to define a distance between the networks. The effectiveness of the method is evaluated using a standard classification test, i.e., recognizing the graphs originating from the same synthetic model. Overall, the proposed distances have performances comparable to the best state-of-the-art techniques (graphlet-based methods) with similar computational requirements. Given its simplicity and flexibility, the method is proposed as a viable approach for network comparison tasks.}
\end{abstract}

\begin{document}

\flushbottom
\maketitle

\thispagestyle{empty}



Comparing networks, i.e., quantifying the similarities or differences between graphs, has become a problem of fundamental importance given the impressive growth of available data and, consequently, of network models used. Relevant examples can be found in virtually every application area, from biology to economics, from transportation systems to social media (see, e.g., Refs.\cite{Pr07,van:10,Ali:14,SoEl:14,PiPi:20,brain:20,TaAi:21} for a small sample of the literature). When analyzing a set of networks, the typical tasks are, for example, clustering, i.e., the partitioning of the set to identify groups with similar characteristics; or detecting pattern discontinuities, when time-stamped data (i.e., temporal networks) are available.

To compare two networks, a measure of their dissimilarity has to be defined. A huge number of different approaches have been proposed, some with general applicability and others related to specific domains. They are all aimed at finding a valid compromise between effectiveness, interpretability, and computational efficiency (see Refs.\cite{EmDe:16,DoHo:18,Ta19} for surveys). In this paper, we propose an approach to quantify the dissimilarity between two networks, which falls under the domain of alignment-free methods. No matching between nodes is necessary and practically any pair of graphs (even with different sizes, densities, or from different fields of application) can be compared. In this way, we try to capture the difference in the global structure, rather than the discrepancies in the neighborhood of each specific node.

The simplest and most direct approach to compare two graphs without alignment is to compare global (scalar) network indicators such as clustering coefficient, diameter, or average path length \cite{Ya2014,Ya2015,Faisal:17}. Of course, similar values of global statistics do not necessarily imply similar network structures (e.g., Ref.\cite{Pr07}), and indeed the performance of this approach becomes poor when networks with subtle differences have to be discriminated. Other approaches are based on comparing more complex global network features. For example, spectral methods define a distance between two graphs based on the dissimilarity of the spectra of their adjacency or Laplacian matrix \cite{Wi:08,Gera2018}. The Portrait Divergence method defines a distance by comparing the "portrait" of the two graphs, i.e., a matrix encoding the distribution of shortest-path lengths \cite{Bagrow2019}.

The class of alignment-free methods that generally provides the best performance in classification tasks (i.e., recognizing networks generated by the same model) is the one based on graphlets, which are small induced subgraphs (typically no larger than 5 nodes, for limiting the computational effort). Graphlet-based network distances are based on counting graphlets: the most advanced methods take into account the automorphism orbits \cite{Ya2014,Ya2015}, i.e., differentiate the roles of the nodes in each graphlet. The counts of the graphlets can be organized in several ways: the Graphlet Correlation Distance (GCD) is based on the comparison of the Graphlet Correlation Matrices of the two graphs, which summarize the distribution of the different types of graphlets in the network and the roles played by nodes in each graphlet. Other graphlet-based methods are NetDis \cite{Ali:14}, in which graphlets are counted in 2-step ego-networks, rather than in the whole network, and Grafene \cite{Faisal:17}, which applies principal component analysis to graphlet counts to improve efficiency.

In this paper, we propose an alignment-free method that is based on the distribution, in the network, of a few indicators that locally describe the neighborhood of the nodes. By neighborhood we mean the 1-step ego-network, or \textit{egonet} (hence the name EgoDist, abbreviation for ego-distances, for the distances proposed here), while the indicators we consider are the (normalized) degree, the clustering coefficient, and the egonet persistence (see the next Section for definitions), which describe in increasing detail the connectivity patterns inside and outside the egonet. The statistics of the three indicators can be used, in different combinations, to define a distribution function in 1, 2, or 3 dimensions, which is taken as a synthesis of the network properties. The distance between two networks is then defined as the distance between the corresponding distributions. Thus, similarly to graphlet-based techniques, the local graph structure around each node is summarized by a low-dimensional vector of features, whose statistics are then used as a global network descriptor. However, the features we use are pretty basic and lend themselves to quick computation.

The effectiveness of the proposed ego-distances is evaluated by means of a standard clustering exercise \cite{Ya2014,Ya2015}: a large number of test networks are synthetically generated with seven randomized models, parameterized with different sizes and densities. The proposed ego-distances are then used for classification, that is, to discriminate pairs of graphs originating from the same model from those originating from different models, with size and density acting as confounding factors. The results of the experiments show that the proposed ego-distances perform comparably to the best graphlet-based distance ({Graphlet Correlation Distance} GCD11) -- with similar computational requirements -- despite the challenging experimental environment created by the subtle topological differences between the network models in the pool.


\section*{\label{sec:distance}Methods: Network distance}

Let us consider a network of size $N$ (number of nodes) undirected and unweighted, therefore completely described by the $N\times N$ adjacency matrix $A$, with $A_{ij}=1$ if $i$ and $j$ are connected by an edge, and $A_{ij}=0$ otherwise. If we denote by $L=\frac{1}{2}\sum_{i,j=1}^{N}A_{ij}$ the number of edges, then the density of the network is given by $\rho=\frac{2L}{N(N-1)}$. Given node $i$, its degree $m_i=\sum_{j=1}^{N} A_{ij}$ is the number of neighbors, and its egonet is the induced graph identified by the node set $E_i=\{i\}\cup \{j|A_{ij}=1\}$, i.e., the union of node $i$ and all its neighbors, for a total of $|E_i|=m_i+1$ nodes.

Three scalar quantities, all taking values in $\left[0,1\right]$, can be used to characterize the egonet $E_i$: the \textit{normalized degree}, the \textit{clustering coefficient}, and the \textit{egonet persistence}. 

\subsection*{\label{sec:dd}Normalized degree}

\begin{figure}
	\centering
	\includegraphics[width=17.5cm]{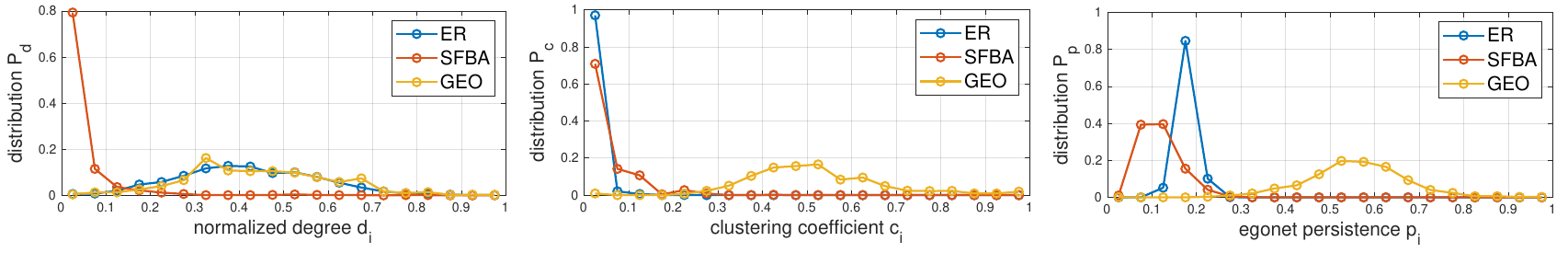}
	\caption{\label{fig:distributions} Distribution of normalized degree (left), clustering coefficient (center), and egonet persistence (right) for three networks of different type: Erd\H{o}s-R\'{e}nyi (ER), Barabasi-Albert (SFBA), and Geometric (GEO) ($N=1000$, $\rho=0.01$, $\Delta=0.05$).}
\end{figure}

For a graph with node degrees ranging in $m_{\min}\le m_i\le m_{\max}$, we define the normalized degree $d_i$ as
\begin{equation}
	d_i=\frac{m_i-m_{\min}}{m_{\max}-m_{\min}},
\end{equation}
where we have assumed $m_{\min}\neq m_{\max}$. We compute the normalized \textit{degree distribution} by discretizing the interval $0\le d_i \le 1$ with step $\Delta$ (then $r=1/\Delta$ is the number of intervals) and by directly calculating the discrete distribution function $P_{d}(h)$ (i.e., the normalized histogram) by counting the proportion of $d_i$'s in each interval. Using the indicator function ($\mathbbm{1}_S x=1$ if $x\in S$ and zero otherwise), we can write:
\begin{equation}\label{eq:Pdd}
	P_{d}(h)=\frac{1}{N}\sum_{i=1}^{N} \mathbbm{1}_{[(h-1)\Delta,h\Delta)}d_i,\quad h=1,2,\ldots,r,
\end{equation}
with values $d_i=1$ conventionally counted in the last interval $h=r$. Figure \ref{fig:distributions} (left panel) shows examples of $P_{d}$ distributions for networks generated by three different -- and well known -- models, with the same size $N$ and density $\rho$. It is preferable to use the cumulative distribution function (cdf) $Q_{d}(h)=\sum_{k=1}^{h}P_{d}(k)$, which is numerically more stable for small $N$. Given the two graphs $G'$ and $G''$, we can define the distance between them as the (Euclidean) distance between the two respective cdf's $Q'_{d}$ and $Q''_{d}$:
\begin{equation}\label{eq:dd}
	D_{d}(G',G'')=\left[\sum_{h=1}^{r}\left(Q'_{d}(h)-Q''_{d}(h)\right)^2\right]^\frac{1}{2}.
\end{equation}
Notice that the above quantity -- as well as the analogous ones defined in the following -- is well defined even when $G'$ and $G''$ have different size, provided that the cdf's are calculated with the same step $\Delta$.

\subsection*{\label{sec:cc}Clustering coefficient}

The (local) clustering coefficient is defined, for a node $i$ with $m_i>1$ neighbors connected by $e_i$ edges, by
\begin{equation}\label{eq:cc}
	c_i=\frac{2e_i}{m_i(m_i-1)},
\end{equation}
while we set $c_i=0$ if $m_i\le 1$. The global (i.e., averaged across all nodes) clustering coefficient $C=\frac{1}{N}\sum_{i=1}^{N} c_i$ was used as a metric to compare networks: the distance between two graphs with clustering coefficients $C'$ and $C''$ is simply given by
\begin{equation}\label{eq:Cglobal}
	D_{C_{global}}=|C'-C''|.
\end{equation} 
The simplicity of this approach makes it very attractive: its performance -- for example, in classification tasks -- is quite good and, in any case, it is the best when compared with the use of other ''simple'' network indicators \cite{Ya2014,Ya2015}.

Here we want to demonstrate that the performance in using the clustering coefficient can be improved by exploiting the distribution of $c_i$'s over the entire network, which is obviously much more informative than just the global value $C$. As we did above for $d_i$, we discretize the interval $0\le c_i \le 1$ and compute the distribution function $P_{c}(h)$ (i.e., the normalized histogram) by counting the proportion of $c_i$'s in each interval. Figure \ref{fig:distributions} (central panel) shows that $P_{c}(h)$ has a good ability to differentiate the three analyzed networks: ER has practically all zero values for $c_i$'s, SFBA has a small number of nonzero (but low) values, while the pdf of GEO has its mass concentrated around $0.5$. Given two graphs $G'$ and $G''$, we can define their distance as:
\begin{equation}\label{eq:dc}
	D_{c}(G',G'')=\left[\sum_{h=1}^{r}\left(Q'_{c}(h)-Q''_{c}(h)\right)^2\right]^\frac{1}{2},
\end{equation}
that is, as the (Euclidean) distance between the corresponding cdf's $Q_{c}(h)=\sum_{k=1}^{h}P_{c}(k)$.

\subsection*{\label{sec:ep}Egonet persistence}

The egonet persistence $p_i$ is defined as the persistence probability \cite{Pi11,De13} of the egonet $E_i$, i.e., the probability that a random walker, located in any of the nodes of $E_i$ at step $t$, remains in any node of $E_i$ at step $t+1$. For an undirected and unweighted network, it can be proved \cite{Pi11} that this quantity is equal to
\begin{equation}\label{eq:p}
	p_i=\frac{\sum_{j\in E_i}m_j^{int}}{\sum_{j\in E_i}m_j} {=\frac{\sum_{j\in E_i}m_j^{int}}{\sum_{j\in E_i}(m_j^{int}+m_j^{ext})}},
\end{equation}
where $m_j^{int}$ {(resp. $m_j^{ext}$)} denotes the internal {(resp. external)} degree of node $j$, i.e., the number of its neighbors internal {(resp. external)} to $E_i$ (we conventionally set $p_i=0$ when $E_i=\{i\}$, i.e., $i$ is an isolated node). As above, we discretize the interval $0\le p_i \le 1$ and we compute the distribution function $P_{p}(h)$ by counting the proportion of $p_i$'s in each interval. 

\begin{figure}[t]
	\centering
	\includegraphics[width=10cm]{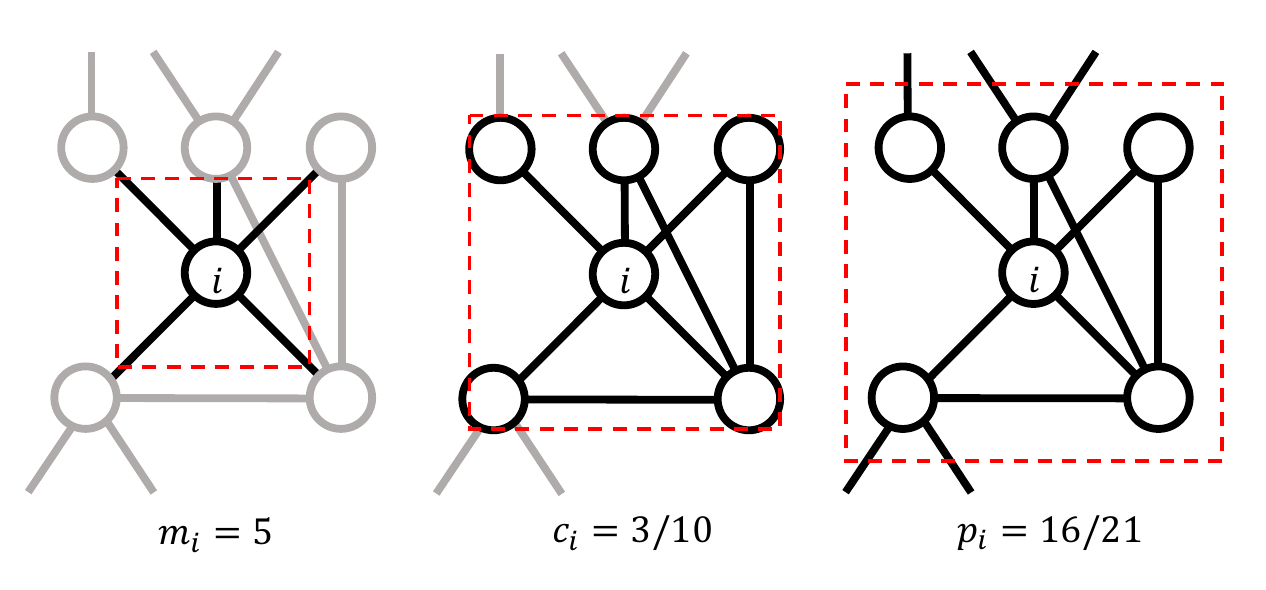}
	\caption{\label{fig:egonets} In the egonet $E_i$, the degree $m_i$ depends only on the connectivity of node $i$ (left); the clustering coefficient $c_i$ describes the connectivity between the neighbors of $i$ (center); the egonet persistence $p_i$ captures the balance between internal and external connectivity of $E_i$ (right){, as it quantifies the proportion of edges that the nodes of the egonet direct into the egonet itself.}}
\end{figure}

\begin{figure}[t]
	\centering
	\includegraphics[width=17.5cm]{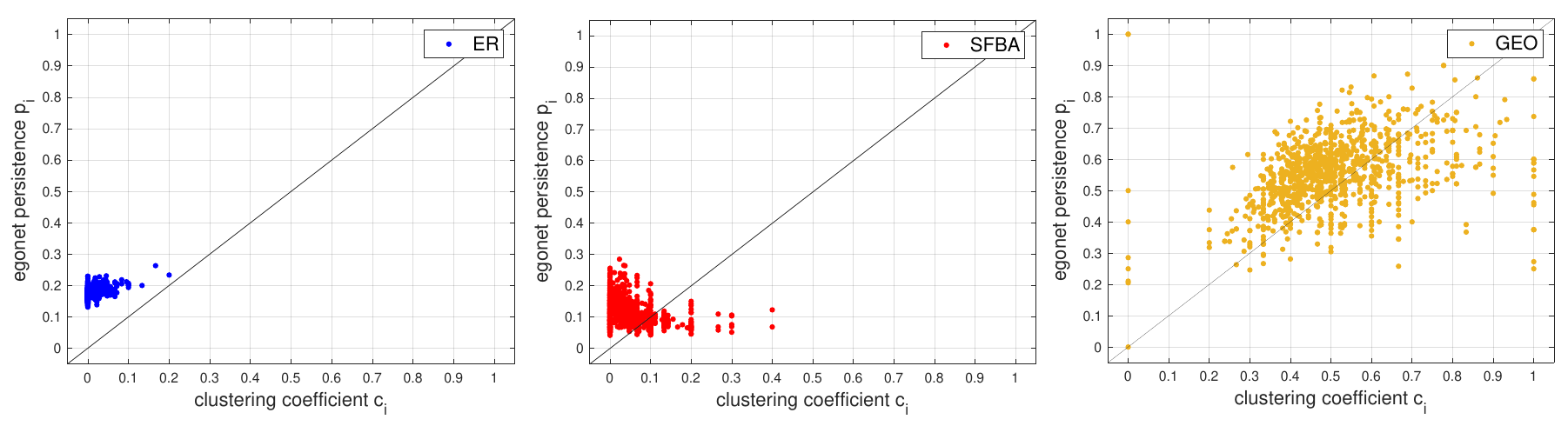}
	\caption{\label{fig:scatter} Scatter plots of clustering coefficient and egonet persistence for the three networks in Fig. \ref{fig:distributions}. The x,y-coordinates of each point are the features of a network node. }
\end{figure}

While the clustering coefficient $c_i$ only captures the connectivity within the egonet, $p_i$ measures the balance between internal and external connectivity, as exemplified in Fig. \ref{fig:egonets}{, i.e., it quantifies the proportion of edges that the nodes of the egonet direct into the egonet itself, rather than to external nodes.} Therefore the two indicators, in general, carry independent information, as emerges from Fig. \ref{fig:distributions}, which shows three examples of the distribution $P_{p}(h)$ for networks generated by different models, and is further demonstrated in Fig. \ref{fig:scatter}, where 2-dimensional scatter plots are shown for the same three networks.

Given the two graphs $G'$ and $G''$, we define their distance as:
\begin{equation}\label{eq:dp}
	D_{p}(G',G'')=\left[\sum_{h=1}^{r}\left(Q'_{p}(h)-Q''_{p}(h)\right)^2\right]^\frac{1}{2},
\end{equation}
i.e., as the (Euclidean) distance between the corresponding cdf's $Q_{p}(h)=\sum_{k=1}^{h}P_{p}(k)$.

\subsection*{\label{sec:combo}Multi-dimensional distributions and distances}

So far we introduced three graph-to-graph distances, respectively $D_{d}$, $D_{c}$, and $D_{p}$, each based on a single indicator. Since normalized degree, clustering coefficient, and egonet persistence carry in general independent information, it is obviously possible to combine them to define more complex metrics. A simple solution would be to add them:
\begin{equation}\label{eq:sum}
	D_{sum}(G',G'')=D_{d}(G',G'')+	D_{c}(G',G'')+D_{p}(G',G''),
\end{equation}
or, more in general, to combine them linearly (this would require, however, properly adjusting the parameters of the linear combination). 

A more general approach is inspired by the plots in Fig. \ref{fig:scatter}, which clearly show that different network models give rise to different distributions of the two-dimensional variable $(c_i,p_i)$. Therefore, we define the two-dimensional discrete distribution function $P_{c,p}(h,k)$ as the normalized 2D histogram:
\begin{equation}\label{eq:P2D}
	P_{c,p}(h,k)=\frac{1}{N}\sum_{i=1}^{N} (\mathbbm{1}_{[(h-1)\Delta,h\Delta)}c_i \times \mathbbm{1}_{[(k-1)\Delta,k\Delta)}p_i),\quad
	h,k=1,2,\ldots,r,
\end{equation}
with values $c_i=1$ (resp. $p_i=1$) conventionally counted in the last interval $h=r$ (resp. $k=r$). Given the two graphs $G'$ and $G''$, we define their distance as 
\begin{equation}\label{eq:dcp}
	D_{c,p}(G',G'')=\left[\sum_{h,k=1}^{r}\left(Q'_{c,p}(h,k)-Q''_{c,p}(h,k)\right)^2\right]^\frac{1}{2},
\end{equation}
that is, the Frobenius norm of the difference between their cdf's $Q_{c,p}(h,k)=\sum_{i=1}^{h}\sum_{j=1}^{k}P_{c,p}(i,j)$. Of course, in the same way as we did in \eqref{eq:P2D}, \eqref{eq:dcp}, we can define two more 2D measures, namely $D_{d,c}(G',G'')$, based on the two-dimensional variable $(d_i,c_i)$, and $D_{d,p}(G',G'')$, based on the pair $(d_i,p_i)$.

\begin{figure}
	\centering
	\includegraphics[width=17.5cm]{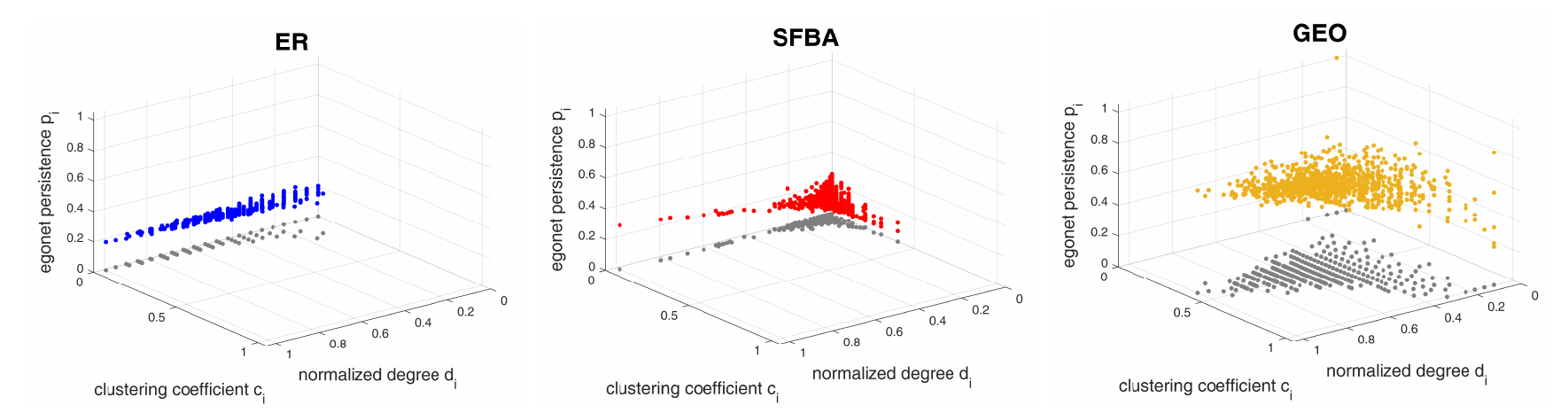}
	\caption{\label{fig:scatter3d} Scatter plots of normalized degree, clustering coefficient, and egonet persistence for the three networks of Fig. \ref{fig:distributions}. The x,y,z-coordinates of each point {(colored dots)} are the egonet features of a network node. {The gray dots are the projection onto the horizontal plane and are shown for readability only.} }
\end{figure}

Finally, a distance measure that fully exploits all the available information is obtained by considering the multivariate distribution of the three-dimensional variable $(d_i,c_i,p_i)$, which captures the different patterns originating from the different network models (Fig. \ref{fig:scatter3d}). This requires partitioning the set $[0,1]^3$ into $r^3$ discretization cubes, computing the three-dimensional discrete distribution function $P_{d,c,p}(h,k,n)$ as:
\begin{equation}\label{eq:P3D}
	P_{d,c,p}(h,k,n)=\frac{1}{N}\sum_{i=1}^{N}  ( \mathbbm{1}_{[(h-1)\Delta,h\Delta)}d_i \times  
	\mathbbm{1}_{[(k-1)\Delta,k\Delta)}c_i \times \mathbbm{1}_{[(n-1)\Delta,n\Delta)}p_i) ,\quad
	h,k,n=1,2,\ldots,r,
\end{equation}
and defining the network distance between $G'$ and $G''$ as
\begin{equation}\label{eq:d3p}
	D_{d,c,p}(G',G'')=
	\left[\sum_{h,k,n=1}^{r}\left(Q'_{d,c,p}(h,k,n)-Q''_{d,c,p}(h,k,n)\right)^2\right]^\frac{1}{2},
\end{equation}
where $Q_{d,c,p}(h,k,n)=\sum_{i=1}^{h}\sum_{j=1}^{k}\sum_{l=1}^{n}P_{d,c,p}(i,j,l)$ is the cdf.

In the following, we will generically indicate with \textit{ego-distances} the graph-to-graph measures introduced in this section and based on the statistics of the egonet indicators: $D_d$, $D_c$, $D_p$, $D_{sum}$ (1D distances), $D_{c,p}$, $D_{d,c}$, $D_{d,p}$ (2D), and $D_{d,c,p}$ (3D).

\section*{\label{sec:evaluation} Results}

\subsection*{\label{sec:classification}Classification of synthetic networks}

To have a fair evaluation of the efficacy of the ego-distances proposed above, we reproduce the experimental setup described in Refs. \cite{Ya2014,Ya2015} by generating synthetic networks from the same seven models used therein (see the section \textit{Materials} for details and references): Erd\H{o}s-R\'{e}nyi model (ER); ER degree distribution preserving model (ERDD); Barab\'asi-Albert scale-free preferential attachment model (SFBA); scale-free gene duplication and divergence model (SFGD); geometric random graph model (GEO); geometric model with gene duplication (GEOGD); stickiness-index based model (STICKY). For each model, we generate networks with size $N=1000$, $2000$, and $4000$, and density $\rho=0.004$, $0.01$, and $0.02$, for a total of $7\times 3\times 3=63$ combinations model/size/density. For each combination, we randomly generate $10$ network instances, so that the experimental setup includes $630$ networks.

{It should be noted that most of the above network models do not necessarily produce a connected network (SFBA and ERDD are the only exceptions - see the section \textit{Materials} for details). For example, at $\rho=0.004$ most of the networks in our sample are not connected. However, the proposed measures present no problems in managing non connected networks. In particular, the three proposed egonet features are well defined also for isolated nodes: they all assume zero value, thus shaping the distributions of the features in such a way as to suitably characterize the network.}

For each pair of networks, we compute the ego-distances defined above, using discretization step $\Delta=0.01$ (the results are largely insensitive to this parameter thanks to the use of cdf's). We also consider $D_{C_{global}}$ (eq. (\ref{eq:Cglobal})), which is an easy-to-calculate distance based on a global network feature. Finally, to have a challenging comparison, we compute the Graphlet Correlation Distance GCD11 ($D_{GCD11}$) \cite{Ya2014,Ya2015}, which is considered one of the most effective distances \cite{Ta19}. 

The goal of the classification task is to recognize when two networks come from the same model. For this purpose, the performance of each distance is evaluated in the usual Precision/Recall framework: two networks form an \textit{actual positive} pair if they are generated by the same model, an  \textit{actual negative} pair otherwise. To be effective for correct classification, the distance between two networks generated by the same model should be much smaller than the distance between two networks originating from different models. Given a distance $D$ and a threshold $\varepsilon>0$, a network pair is a  \textit{predicted positive} sample if $D<\varepsilon$, a  \textit{predicted negative} sample otherwise. Then Precision and Recall are given, for each $\varepsilon$, by $P_\varepsilon=tp/(tp+fp)$ and $R_\varepsilon=tp/(tp+fn)$, where \textit{tp}, \textit{fp}, and \textit{fn} are, respectively, the number of \textit{true positive}, \textit{false positive}, and \textit{false negative} network pairs. The Precision/Recall curve provides a graphical representation of the simultaneous evolution of $P$ and $R$ with $\varepsilon$, and the area under the curve ({denoted by AUPR, i.e., \textit{Area Under the Precision/Recall curve}}, $0\le \text{AUPR}\le 1$) is a quantity that summarizes the performance of each distance, with the limit $\text{AUPR}=1$ obtained in the ideal case\cite{Davis2006,Saito2015}.

As clearly highlighted in Ref.\cite{Ya2015}, an ideal distance should be able to recognize networks generated by the same model without being confused by possible differences in size and density, but only being influenced by structural differences in the topology -- a very challenging task. On the other hand, a distance can be considered good, though not ideal, when it is able to perform the above task at least when the networks are of the same size and density. For this reason, we also systematically evaluate the performance of each distance on a subset of network pairs, namely those (possibly) generated by different models but having the same size and density.

\begin{figure}
	\centering
	\includegraphics[width=7cm]{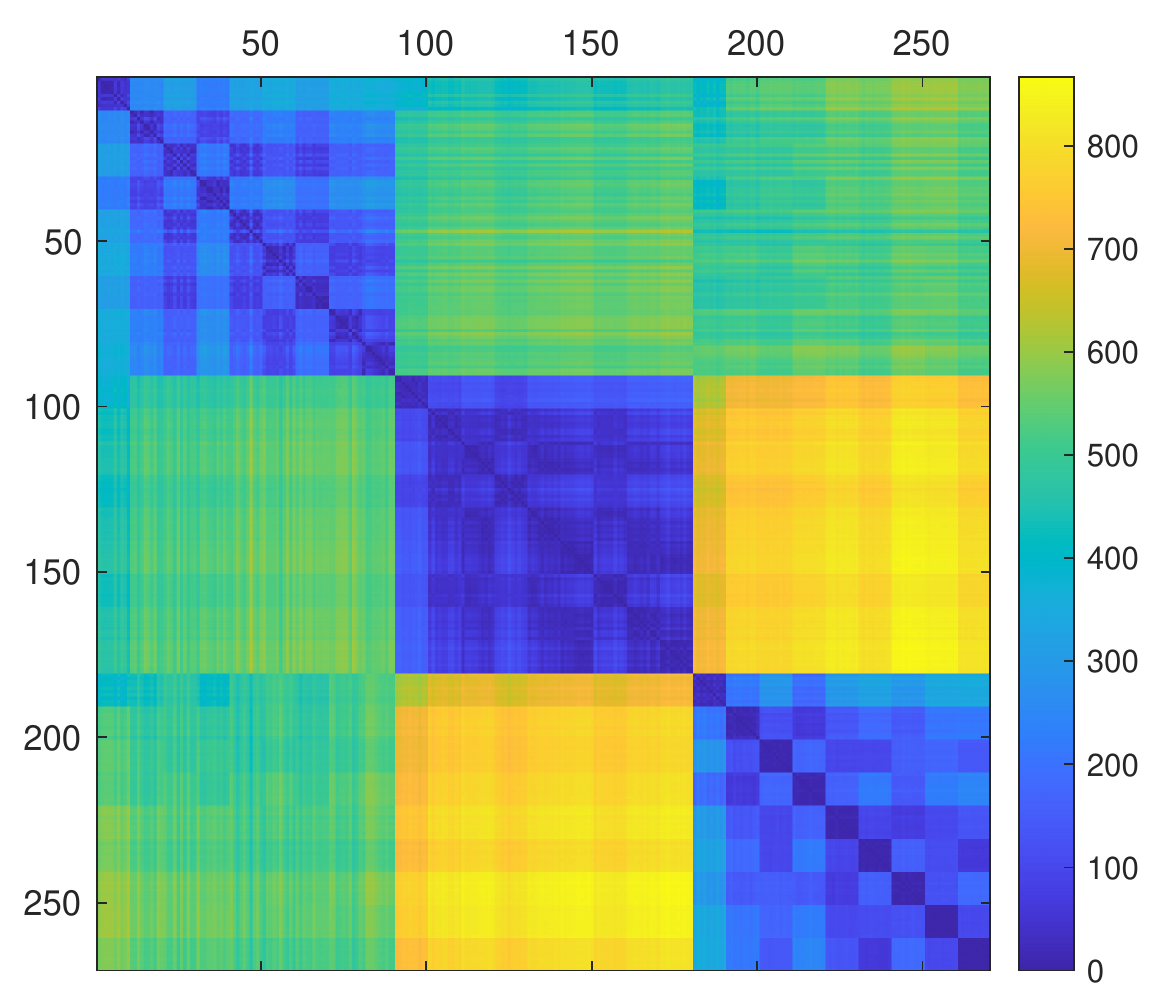}
	\caption{\label{fig:d_esg} Distance matrix $D_{d,c,p}$ between the networks ER, GEO, and SFBA (3 models $\times$ 3 sizes $\times$ 3 densities $\times$ 10 replications = 270 networks): the different models are clearly separated, regardless of the different combinations of size/distances.}
\end{figure}

We start by limiting the analysis to just three network models, namely ER, GEO, and SFBA. In the previous section, their characteristics were compared in Figs. \ref{fig:distributions}, \ref{fig:scatter}, and \ref{fig:scatter3d}, showing that each model has some peculiarities -- in terms of the adopted egonet features -- which differentiate it from the other two. Thus we expect that the proposed ego-distances are able to correctly classify the network model: Table \ref{tab:esg} shows that this is indeed the case for all ego-distances if size and density are the same, but also the same happens for most ego-distances when networks are of different size and/or density. In other words, for this simplified task the egonet features are sufficiently differentiated that most combinations of two or three cdf's will distinguish the three models perfectly. As an example, we report in Fig. \ref{fig:d_esg} the distance matrix relating to $D_{d,c,p}$, which highlights the clear separation between the different models.

\begin{table}
	\centering
	\caption{\label{tab:esg} AUPR {(Area Under the Precision/Recall curve)} value for the classification of ER, GEO, and SFBA networks, for the ego-distances defined in the section \textit{Methods} and for the {Graphlet Correlation Distance} GCD11 distance \cite{Ya2014}. The best-ranked distances are highlighted in bold italic. }
		\begin{tabular}{ccc}
			\toprule
			{distance}&{all sizes/densities}&{same size/density}\\
			D & 0.781 & 0.873 \\
			C & 0.808 & 0.997 \\
			P & 0.777 & \textit{\textbf{1}} \\
			SUM & \textit{\textbf{0.993}} & \textit{\textbf{1}} \\
			C,P & 0.827 & \textit{\textbf{1}}\\
			D,C & \textbf{\textit{1}} & \textit{\textbf{1}}\\
			D,P & \textit{\textbf{0.993}} & \textit{\textbf{1}}\\
			D,C,P & \textbf{\textit{0.999}} & \textit{\textbf{1}}\\
			\midrule
			$\text{C}_{global}$ & 0.782 & 0.994\\			
			GCD11 & 0.649 & 0.995\\	
			\bottomrule		
		\end{tabular}
\end{table}

Much more challenging is the task of recognizing the correct model when there are subtle topological differences, as in the case where all seven models are in the pool: the results are summarized in Table \ref{tab:all} (upper part). In terms of AUPR, all the proposed ego-distances outperform GDC11 when sizes/densities are mixed. On the other hand, GCD11 achieves much better results in the simpler task of coupling networks of the same size/density.

\begin{table}
	\centering
	\caption{\label{tab:all}AUPR {(Area Under the Precision/Recall curve)} value for the classification of all network models, for the distances defined in the section \textit{Methods} and for the {Graphlet Correlation Distance} GCD11 distance \cite{Ya2014}. The best-ranked distances are highlighted in bold italic.}
		\begin{tabular}{ccc}
			\toprule
			\multicolumn{3}{c}{\textbf{cap value} $T=1$} \\
			\midrule
			{distance}&{all sizes/densities}&{same size/density}\\
			D & 0.542 & 0.579 \\
			C & 0.436 & 0.609 \\
			P & 0.379 & 0.631 \\
			SUM & 0.564 & 0.666 \\
			C,P & 0.457 & 0.669 \\
			D,C & \textit{\textbf{0.631}} & 0.653\\
			D,P & \textit{\textbf{0.593}} & \textbf{\textit{0.673}}\\
			D,C,P & \textit{\textbf{0.613}} & \textbf{\textit{0.702}}\\
			\toprule
			\multicolumn{3}{c}{\textbf{cap value} $T=0.5$} \\
			\midrule
			{distance}&{all sizes/densities}&{same size/density}\\
			D & 0.554 & 0.581 \\
			C & 0.442 & 0.646 \\
			P & 0.450 & 0.693 \\
			SUM & \textit{\textbf{0.596}} & 0.692 \\
			C,P & 0.496 & 0.774 \\
			D,C & \textit{\textbf{0.628}} & 0.743\\
			D,P & \textbf{\textit{0.617}} & \textit{\textbf{0.775}}\\
			D,C,P & 0.571 & \textit{\textbf{0.798}}\\
			\midrule
			$\text{C}_{global}$ & 0.389 & 0.516\\			
			GCD11 & 0.422 & \textbf{\textit{0.863}}\\	
			\bottomrule		
		\end{tabular}
\end{table}

\begin{figure}
	\centering
	\includegraphics[width=8cm]{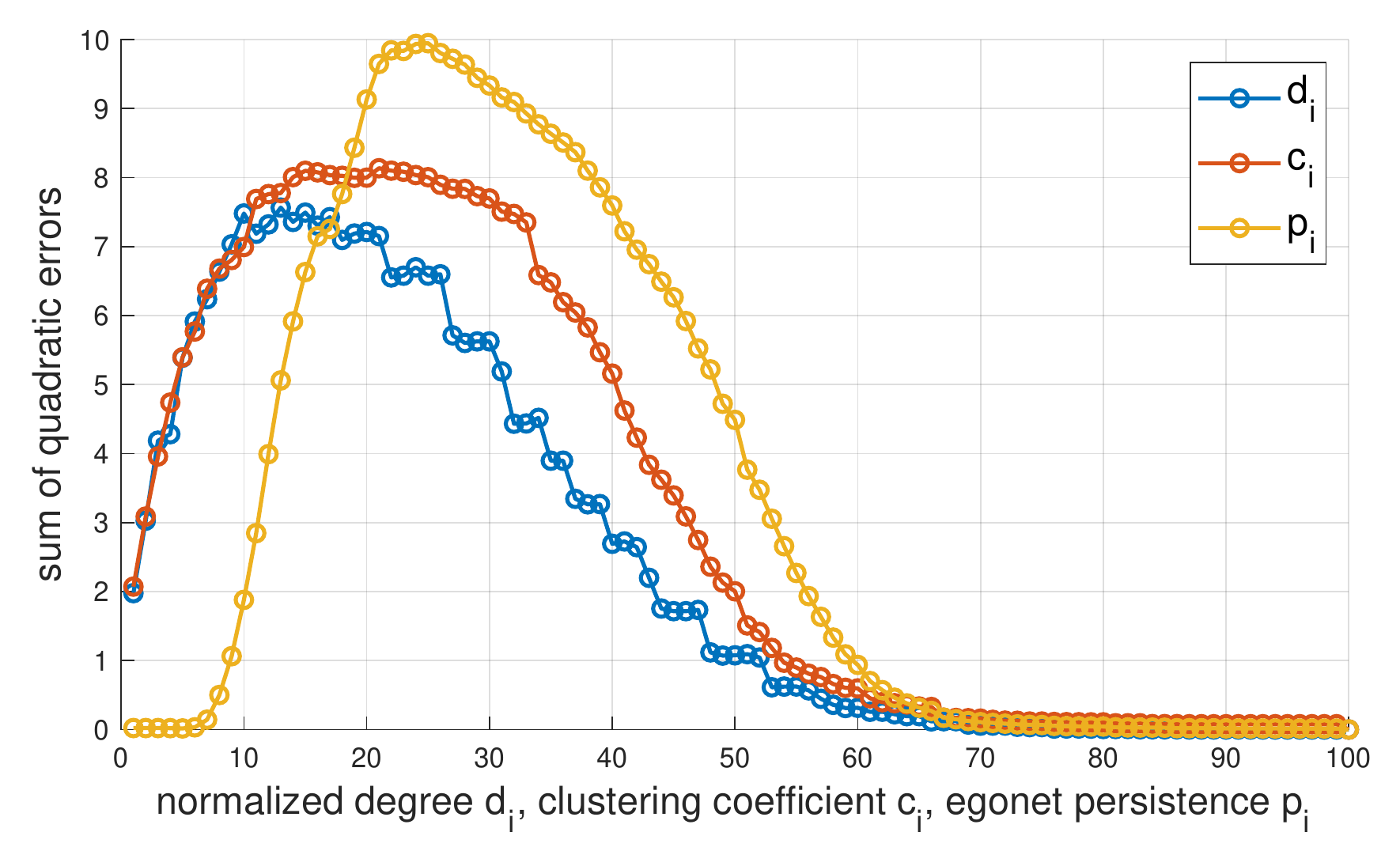}
	\caption{\label{fig:quaderrors} The quadratic errors $\sum (Q'_{*}(h)-Q''_{*}(h))^2$ as a function of the bin index $h=1,2,\ldots,r$ ($r=1/\Delta$, $\Delta=0.01$, $*=d,c,p$). The sum is extended to the pairs of networks formed by pairing 7 networks, one for each model, in all possible combinations ($N=1000$, $\rho=0.01$).    }
\end{figure}

The performance of ego-distances further improves if we restrict the computation of the cdf's of $d_i$, $c_i$, $p_i$ to a sub-range of $[0,1]$. Indeed, if we analyze the quadratic error terms $(Q'_{*}(h)-Q''_{*}(h))^2$ which form the core of the distances \eqref{eq:dd}, \eqref{eq:dc}, \eqref{eq:dp} and draw their dependence on $h=1,2,\ldots,r$ covering the interval $[0,1]$, we see (Fig. \ref{fig:quaderrors}) that the differences between networks vanish as the upper bound is approached -- not surprising, since $d_i$, $c_i$, $p_i$ rarely take values close to $1$. Therefore, by restricting the computation to the range where the above terms are significant, we increase their sensitivity and, as a by-product, reduce the computational effort. For this purpose, we define a cap value $0<T\le 1$ {(i.e., a maximum value)} as the upper limit for computing the cdf differences: in \eqref{eq:dd}, \eqref{eq:dc}, \eqref{eq:dp}, all sums remain extended to $h=1,2,\ldots,r$, but now $r=T/\Delta$ instead of $r=1/\Delta$. 

\begin{figure}
	\centering
	\includegraphics[width=17.5cm]{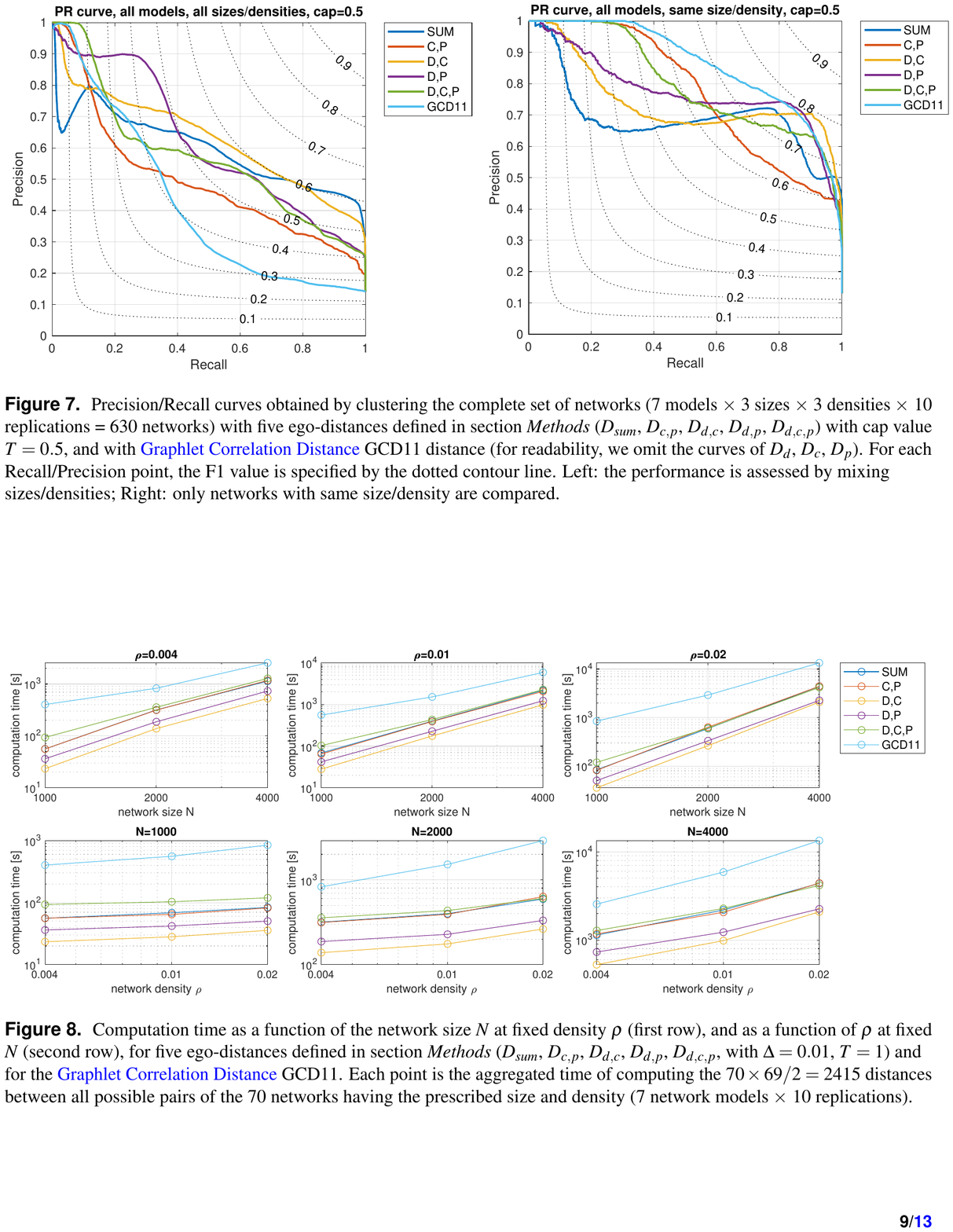}
	\caption{\label{fig:PR_all_cap05} Precision/Recall curves obtained by clustering the complete set of networks (7 models $\times$ 3 sizes $\times$ 3 densities $\times$ 10 replications = 630 networks) with five ego-distances defined in section \textit{Methods} ($D_{sum}$, $D_{c,p}$, $D_{d,c}$, $D_{d,p}$, $D_{d,c,p}$) with cap value $T=0.5$, and with {Graphlet Correlation Distance} GCD11 distance (for readability, we omit the curves of $D_d$, $D_c$, $D_p$). For each Recall/Precision point, the F1 value is specified by the dotted contour line. Left: the performance is assessed by mixing sizes/densities; Right: only networks with same size/density are compared. }
\end{figure}

Table \ref{tab:all} (lower part) and Fig. \ref{fig:PR_all_cap05} show the results obtained with cap value $T=0.5$. Compared to the case $T=1$, the AUPR values do not vary remarkably when all sizes/densities are mixed, while they increase significantly when comparing networks with the same size/density. The Precision/Recall curves of Fig. \ref{fig:PR_all_cap05} (left panel) show that the ego-distances obtain qualitatively similar behavior, with $D_{d,c,p}$ maintaining the largest Precision at very small Recall values, and $D_{d,c}$ and $D_{sum}$ yielding the best compromise at large Recall values, as evidenced by the largest F1 value (we remind that $\text{F1}=2PR/(P+R)$ is the harmonic mean of Precision and Recall). On the other hand, when the comparison is restricted to networks with the same size/density, Fig. \ref{fig:PR_all_cap05} (right panel) confirms that the {Graphlet Correlation Distance} $D_{GCD11}$ remains overall superior, although some ego-distances are able to achieve comparable F1 values.

{We now summarize the above results, in order to provide guidance on which  of the proposed ego-distances to adopt. We observe that our results depend on the experimental dataset, which is rich and diversified but, obviously, does not include all possible network structures: any conclusion must therefore be evaluated with caution. In general, the AUPR (Table \ref{tab:all}) should be the primary selection criterion, as it represents the average precision across all possible recall values. Not surprisingly, AUPR highlights the superiority of 2D and 3D measures, which exploit more information than 1D measures. In most cases, $D_{d,c,p}$ performs best (or nearly so) and should therefore be considered the preferred metric - an expected result, since it uses the most information in the most structured form, that is, building 3D distributions. Furthermore, the Precision/Recall curves of Fig. \ref{fig:PR_all_cap05} show that $D_{d,c,p}$ keeps large Precision for Recall$\rightarrow 0$ (left side of the curve); has a monotone and sufficiently regular behavior; and grants a rather large Precision even for Recall$\rightarrow 1$ (right side of the curve). As we will see in the next section (\textit{Computational requirements}), however, $D_{d,c,p}$ is the most computationally expensive of the ego-distances. If its use is prohibitive in a specific application, $D_{d,c}$, $D_{d,p}$ are valid alternatives, as shown in Table \ref{tab:all}.      }

\subsection*{\label{sec:computation}Computational requirements}

The same pool of synthetic networks, previously used to evaluate the classification capabilities of the proposed ego-distances, was also exploited to empirically test their computational requirements. A theoretically-based prediction is quite difficult, given the mixed sequence of operations involved in each ego-distance, namely the computation of one or more egonet features, of the cdf's, and of their Euclidean distance. The first task (computing the egonet features), however, is definitely dominant for medium to large networks. In general terms, assuming that checking the connection of a node pair $(i,j)$ requires a fixed time, then computing the degree of a node requires time $O(N)$, which becomes $O(N^2)$ for all $N$ nodes. For the clustering coefficient (eq. \eqref{eq:cc}), checking connections between the neighbors of a node of degree $m_i$ requires $O(m_i^2)$ operations, which is $O(N^2)$ is the worst case and therefore $O(N^3)$ for the whole network. Analogously for the egonet persistence (eq. \eqref{eq:p}): in the numerator we check the $(m_i+1)^2$ possible connections internal to $E_i$, while in the denominator the $(m_i+1)N$ possible connections of the nodes of $E_i$ with all nodes of the network: both terms are $O(N^2)$ in the worst case, which leads to a complexity $O(N^3)$ for all $N$ nodes. On the other hand, the computational complexity of {Graphlet Correlation Distance} GCD11 is $O(Nm_{\max}^3)$ \cite{HoDe14,Ya2015}. However, typical networks are often far from the worst case and therefore the computational requirements are milder.

To get an empirical estimate of the computational requirements of the ego-distances, we run experiments for all nine combinations of size ($N=1000$, $2000$, and $4000$) and density ($\rho=0.004$, $0.01$, and $0.02$) used above: for each pair $(N,\rho)$, we compute the $70\times 69/2=2415$ distances between all possible pairs of the $70$ networks having the prescribed size and density (recall that we have $7$ network models and $10$ replicas for each model), and we add up the time required for the computation. Finally, we have an aggregated time for each pair $(N,\rho)$, which is obtained from models with mixed characteristics and thus is representative of the average computational requirements of the distance used.

\begin{figure*}
	\centering
	\includegraphics[width=17.5cm]{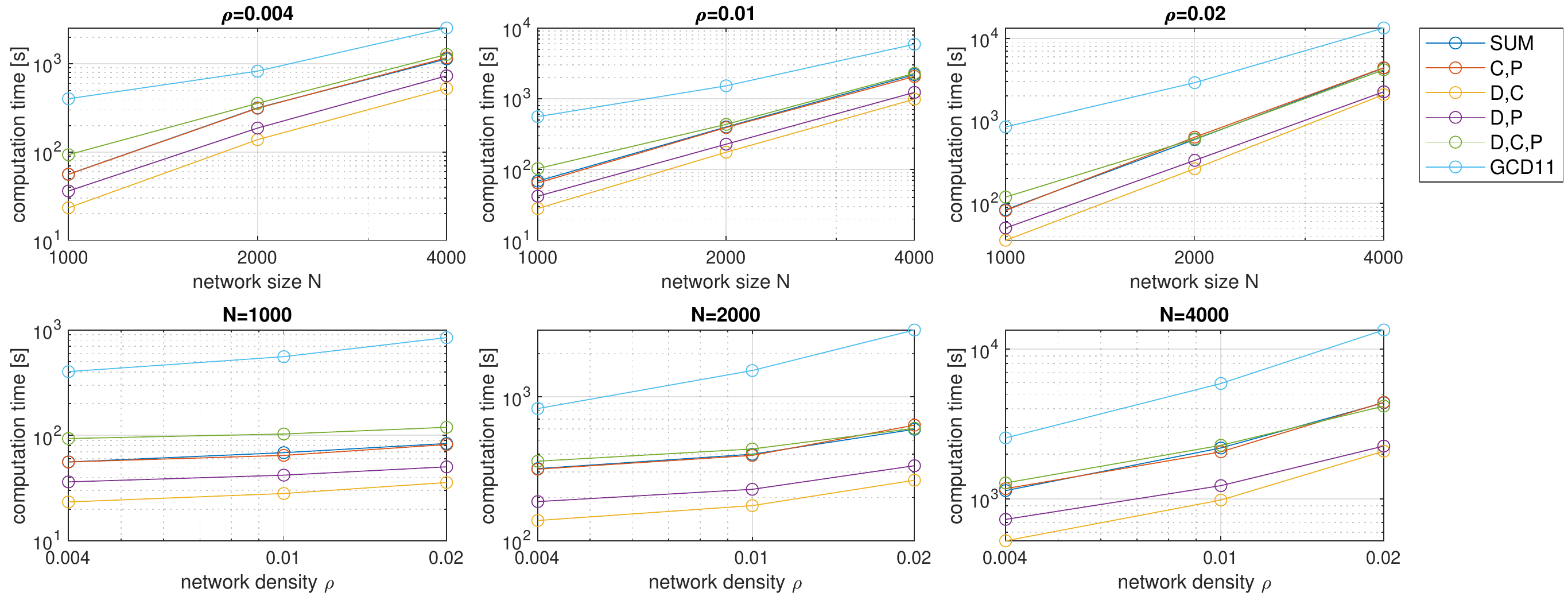}
	\caption{\label{fig:time_N_rho} Computation time as a function of the network size $N$ at fixed density $\rho$ (first row), and as a function of $\rho$ at fixed $N$ (second row), for five ego-distances defined in section \textit{Methods} ($D_{sum}$, $D_{c,p}$, $D_{d,c}$, $D_{d,p}$, $D_{d,c,p}$, with $\Delta=0.01$, $T=1$) and for the {Graphlet Correlation Distance} GCD11. Each point is the aggregated time of computing the $70\times 69/2=2415$ distances between all possible pairs of the $70$ networks having the prescribed size and density ($7$ network models $\times$ $10$ replications). }
\end{figure*}

Figure \ref{fig:time_N_rho} shows the results of the above experiments. First of all, we must mention that the time needed to compute the three ego-distances  $D_d$, $D_c$, and $D_p$ (omitted in the figure) increases from the first to the last, as one would expect, with the first almost negligible compared to the others. This is why it clearly emerges from the figure that, for fixed $N$ and $\rho$, $D_{d,c}$ and $D_{d,p}$ (in that order) are the fastest to compute, followed by $D_{c,p}$, which is based on the two most expensive features, and by $D_{sum}$ and $D_{d,c,p}$, which require all three features (the figure shows that the latter is the slowest of the ego-distances, due to the need to elaborate 3D distributions). All ego-distances are faster than GCD11 anyway (at least in our implementation, see section \textit{Materials} below for details). The three panels in the first row of Fig. \ref{fig:time_N_rho} show, for all $\rho$ values, a computation time approximately increasing as $t\propto N^\alpha$, with $\alpha$ between $1.89$ and $2.93$. GCD11 scales more favorably in this regard, with $\alpha$ ranging from $1.33$ to $1.99$. The plots showing the dependence on $\rho$ (second row of the figure) confirm the ranking between the distances in terms of computation time.

{\subsection*{\label{sec:atn}Example of application: European air transportation network}}

\begin{figure*}
	\centering
	\includegraphics{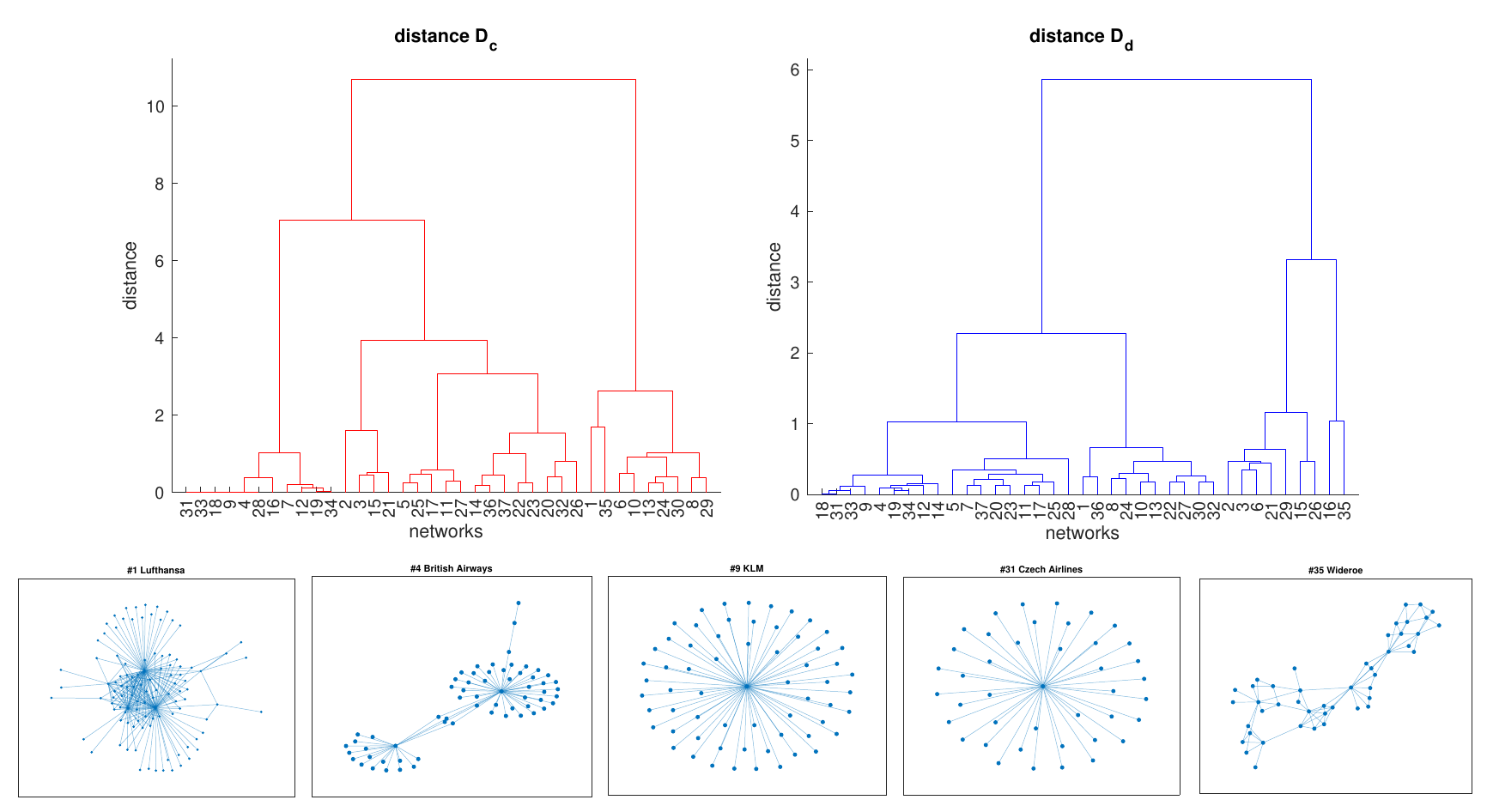}
	\caption{\label{fig:atn} {Top row: Dendrograms of the  hierarchical cluster analysis of the 37 European airline networks, using distances $D_c$ (left) and $D_d$ (right) with average linkage. Bottom row: Five examples of networks from five distinct European airlines. }}
\end{figure*}

{We demonstrate the use of the ego-distances introduced above on data describing the European air transportation network. The dataset includes $37$ networks, corresponding to airlines, each with $ 448 $ nodes (only the connected component of each network is considered, formed by airports connected by flights of the corresponding airline), representing European airports (the complete lists of airports and airlines can be found in Refs. \cite{Cardillo2013,Ta19}). 

To show how ego-distances can be used flexibly to spot specific network (dis)similarities, we compute and compare the results of the two most simplest distances introduced above, namely $D_c$ and $D_d$. Figure \ref{fig:atn} (top row) shows the dendrograms summarizing the results of the hierarchical cluster analysis based on these two distances. In the same figure, the bottom row shows the graphs of five of the networks analyzed, related to five distinct airlines. Despite the apparent dissimilarity of the graphs, the two distances reveal that some of the networks have similar characteristics.}

{For example, networks \#9 (KLM) and \#31 (Czech Airlines) are at zero distance according to $D_c$: they are both pure stars and, as such, all nodes have zero clustering coefficient. But they are also very similar according to $D_d$, as can be read from the right dendrogram in Fig. \ref{fig:atn}. Indeed, the normalized degree distribution is in both cases concentrated on the two extremes $0$ and $1$, corresponding respectively to the minimum and maximum degree, with slightly different values due to the unequal number of nodes in the two cases. 

Despite the different topology, network \#4 (British Airways) also has $D_c=0$ from \#9 and \#31: although the former is not a pure star, it too does not contain triangles and therefore also in this case all nodes have zero clustering coefficient. But network \#4 is quite close to \#9 and \#31 also in terms of $D_d$ because, once again, the degree is concentrated in the extreme values (the leaves and the two hubs), with very few exceptions of nodes with degree $2$.

As a final example, consider networks \#1 (Lufthansa) and \#35 (Wideroe). Although they differ in size and topology, their common trait is the abundance of triangles resulting in similar distributions of the clustering coefficient. Indeed, the dendrogram based on $D_c$ places the two networks at a short distance - and at the maximum possible distance from the pure stars \#4, \#9, \#31. On the other hand, even if similar in terms of $D_c$, the dendrogram based on $D_d$ reveals that networks \#1 and \#35 are the furthest possible, in the analyzed dataset, in terms of degree distribution, a result that could be guessed, to some extent, by looking at the two graphs.}

\vspace{1cm}

\subsection*{\label{sec:realworld}Example of application: Mobility networks during COVID-19 lockdown}

When a sequence of time-stamped networks is available, a typical task is to quantify the (dis)similarity between the graphs to identify anomalous instants in their time evolution. Below we briefly show the results of the analysis of the sequence of mobility networks between Italian cities, estimated from the digital traces of over 4 million individuals \cite{Bonaccorsi21} across the 2020 lockdown period caused by the COVID-19 pandemic.

The dataset includes 32 networks, each of which aggregates the mobility of individuals over a week. The first week begins on February 24, when the first infected individuals had just been detected in Northern Italy and only restrictions on local mobility around the affected cities were imposed. The full lockdown started on March 9 (\url{ https://en.wikipedia.org/wiki/COVID-19_lockdowns_in_Italy}) and has been gradually lifted since May 4. The period covered by the dataset ends in early October. Each network describes the flow of individuals among approximately $3000$ municipalities. We binarize the networks by neglecting weights, thus preserving only the structure, with the aim of analyzing the evolution of the mobility backbone over time. Recall that ego-distances are alignment-free measures, so the focus is on the evolution of the network structure, rather than on the variation of specific flows from city to city.

\begin{figure*}
	\centering
	\includegraphics[width=16cm]{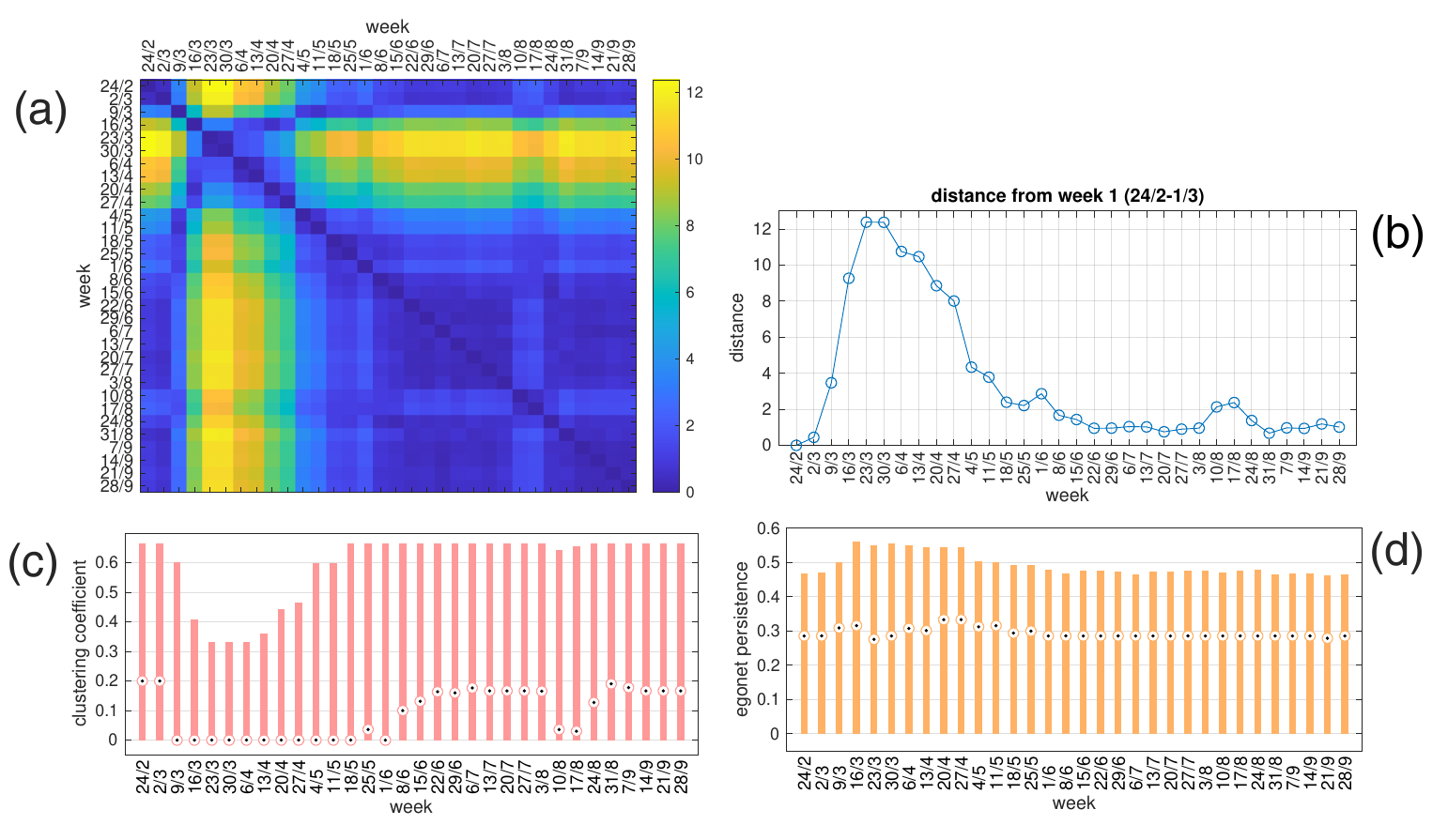}
	\caption{\label{fig:FBmatrix} (a) The $D_{c,p}$ distance matrix among the 32 Italian weekly mobility networks during  the COVID-19 lockdown period in 2020: the date on the axes is the first day of the corresponding week. (b) The first row of the distance matrix $D_{c,p}$, i.e., the distance of each mobility network from the week 1 network. (c)-(d) Boxplots of the distribution of the clustering coefficient (c) and egonet persistence (d). The boxes range from the 25th to the 75th percentile, the point being the median value.}
\end{figure*}

The results using $D_{c,p}$ are summarized in Fig. \ref{fig:FBmatrix} (similar results are obtained with other ego-distances). Panels (a)-(b) clearly highlight the strong anomaly of the network structure during the lockdown period. Interestingly, although the nationwide measures (including the mobility block) had been in place since March 9, the plots show that it took a couple of weeks to reach the maximum deviation from the unperturbed situation (it was only on March 21 that it was decided to close all unnecessary businesses and industries), and that such a regime lasted for a rather short period. In fact, the return to normal regime begins a couple of weeks before May 4, when intra-regional mobility was allowed. After that date, the mobility network returns to be very similar to that of the pre-COVID period. Incidentally, panel (b) reports, during the summer period, small alterations in mobility in the days close to Republic Day (June 2) and during the central weeks of August, the traditional holiday weeks.

Panels (c)-(d) show how the structural variation of the network is captured by the two ego-features included in $D_{c,p}$. Due to the interruption of many mobility corridors, the median of the clustering coefficient drastically drops (remains at zero for many weeks), and the entire distribution is strongly compressed. Less evident is the impact on the egonet persistence, whose median does not vary much: the distribution, however, becomes wider during the weeks of lockdown, because a number of egonets become more cohesive due to the interruption of external connections (see Fig. \ref{fig:egonets}).

\section*{\label{sec:conclusions}Concluding remarks}

This work introduced a family of alignment-free network-to-network dissimilarity measures (EgoDist), based on the comparison of the distributions, on the network, of a few features that locally characterize the egonets: the degree, the clustering coefficient, and the egonet persistence. The dissimilarity between two graphs is defined as the distance between the corresponding distributions (one-dimensional or multi-dimensional). The ability of the proposed measures to discriminate networks with subtly different characteristics was evaluated by means of a standard experimental setup. Overall, EgoDist measures perform comparably to graphlet-based measures, with similar computational requirements.

The method has several possible generalizations. On the one hand, the extension to weighted and/or directed networks is conceptually immediate, since all the egonet features adopted here have their relevant generalizations \cite{Fagiolo2007,Pi11}. On the other hand, the set of node features can be increased by using, for example, the distribution of any centrality indicator of nodes\cite{Ne:10,Latora2017,Wang2018}, instead of, or in addition to, the quantities used here (this  however implies loosening the assumption of using only egonet features). Incidentally, this opens up the problem of finding the (minimum) set of indicators that achieves the "best" network classification.

It goes without saying that all of the above generalizations imply a significant increase in computational requirements. In this regard, to keep the method scalable to large-scale networks one could exploit parallelization techniques (the computation of egonet features can be completely parallelized) but also devise sampling techniques, i.e., compute the distributions of the egonet features on a sample of nodes rather than across the entire network. This obviously requires the use of graph sampling techniques \cite{Ahmed13}, whose effect on the performance of the proposed approach should be thoroughly evaluated.


\section*{Materials}

\subsection*{Network models}

Below we summarize the description of the algorithms for the generation of the seven synthetic network models used for the evaluation experiments: to have comparable results, they are the same used by \cite{Ya2014,Ya2015}. In all cases, the network is defined by the size $N$ (number of nodes) and the density $\rho=\frac{2L}{N(N-1)}$, where $L$ is number of edges. Notice that the average degree $m_{avg}=\frac{2L}{N}$ can be expressed as $m_{avg}=\rho(N-1)$.

\textbf{ER (Erd\H{o}s-R\'{e}nyi model)} Each node pair $(i,j)$ is connected with probability $\rho$ \cite{ErRe:59,Ne:10}.

\textbf{SFBA (Barab\'asi-Albert scale-free preferential attachment model)} We define $\eta=\frac{\rho N}{2}$, which is $\frac{m_{avg}}{2}$ for large $N$, and note that $\eta$ assumes integer values for all the pairs $(N,\rho)$ used in the article. We initialize the network with a clique (complete graph) of $\eta+1$ nodes, then add one node at a time until we reach the prescribed size $N$. Each added node must connect its $\eta$ edges to $\eta$ target nodes, which are randomly selected with probability proportional to their degree in the current network (preferential attachment \cite{BaAl:99,Ba:16}).

\textbf{ERDD (ER degree distribution preserving model)} An SFBA network is first created with the prescribed $N,\rho$ (see above), then all edges are shuffled while preserving the individual degree of each node (degree-preserving randomization \cite{maslov02,Ne:10}).

\textbf{STICKY (stickiness-index based model)} An SFBA network is first created with the prescribed $N,\rho$ (see above), thus defining the degree sequence ${m_1,m_2,\ldots,m_N}$: then all edges are removed. Finally, each pair of nodes $(i,j)$ is connected with probability $\frac{m_i m_j}{\sum_h m_h}$ \cite{przulj2006}.

\textbf{SFGD (scale-free gene duplication and divergence model)} We initialize the network with a seed of $2$ connected nodes, then add one node at a time until we reach the prescribed size $N$. For each node $i$ to be added, an existing node $j$ is selected uniformly at random, and $i$ is connected to all neighbors of $j$. Furthermore, the pair $(i,j)$ is connected with probability $0.5$. Then, we consider all nodes $h$ that are common neighbors of $i$ and $j$ and, with probability $q$, remove either the edge $(h,i)$ or $(h,j)$ (with random selection). The value of $q$ is iteratively adjusted to reach the prescribed density $\rho$ (on average over the 10 network replications)\cite{vazquez2003}.

\textbf{GEO (geometric random graph model)} The $N$ nodes are thought of as points in the unit cube, whose 3D coordinates are selected uniformly at random. Then the nodes $(i,j)$ are connected if and only if their Euclidean distance is smaller than a given $r$, the value of which is iteratively adjusted to reach the prescribed density $\rho$ (on average over the 10 network replications)\cite{penrose2003}.

\textbf{GEOGD (geometric model with gene duplication)}  The $N$ nodes are thought of as points in the unit cube. Given a prescribed $r>0$, we initialize the set of nodes with $2$ nodes at a much shorter Euclidean distance than $r$, then we add one node at a time until we reach the prescribed size $N$. For each node $i$ to be added, an existing node $j$ is selected uniformly at random, and $i$ is placed in the unit cube at a random position within distance $2r$ from $j$. After all $N$ nodes have been placed, each pair $(i,j)$ is connected by an edge if and only if $(i,j)$ are at smaller Euclidean distance than $r$, the value of which is iteratively adjusted to reach the prescribed density $\rho$ (on average over the 10 network replications)\cite{przulj2010}.

\subsection*{Measuring computation time}\label{app:computation}

The computation times shown in Fig. \ref{fig:time_N_rho} were obtained on a desktop PC with Intel i7 CPU at 2.90GHz using Matlab R2021b. To limit possible confounding factors, the  times reported refer only  to the computation of the distances between the $2415$ network pairs, as described in section \textit{Computational requirements}, i.e., all data loading and organization are ignored. For the EgoDist measurement we used the code available at \url{https://piccardi.faculty.polimi.it/highlights.html}, which implements all the distances proposed in this paper. For {Graphlet Correlation Distance} GCD11 we used the MNA Matlab interface for ORCA \cite{HoDe14}, available at \url{https://github.com/muellsen/MNA/tree/master/GraphletComputation}.

\subsection*{Code and data availability}
The Matlab code of the function EgoDist, implementing the computation of the ego-distances, and the files of the synthetic networks used for the classification task are available at \url{https://piccardi.faculty.polimi.it/highlights.html}. {The data of the European air transportation network are available at \url{http://complex.unizar.es/~atnmultiplex/}.} The mobility network data were kindly provided by the authors of Ref. \cite{Bonaccorsi21}.


\bibliography{egodist}

\section*{Acknowledgements}
The mobility network data were kindly provided by the authors of Ref. \cite{Bonaccorsi21}. The author is grateful to Francesco Pierri for many useful discussions. 

\section*{Author contributions statement}
C.P. conceived the research, conducted the experiments, and wrote the manuscript. 

\section*{Additional information}
The author declares no competing interest.

\end{document}